\documentclass[aps,prl,twocolumn,groupedaddress,fleqn,showkeys,superscriptaddress]{revtex4-1}
\usepackage{chemformula} 
\usepackage[T1]{fontenc} 
\usepackage[english]{babel}
\usepackage{graphicx}
\usepackage{gensymb}
\usepackage{xspace}
\usepackage{fancybox}
\usepackage{epic}
\usepackage{calc}
\usepackage{ifthen}
\usepackage{float}
\usepackage{gensymb}
\usepackage{color}
\usepackage{amsmath,amssymb}
\usepackage{rotating}
\usepackage{subfigure}
\usepackage{exscale}
\usepackage{setspace}
\usepackage{makecell}
\usepackage{epstopdf}
\usepackage{lipsum}
\DeclareMathAlphabet{\mathbfit}{OT1}{cmr}{bx}{it}
\usepackage{multirow}

\begin{document}

\author{S.~Sabet}
\email{sabet@oxide.tu-darmstadt.de}
\affiliation{Institute of Materials Science, Technische Universit\"{a}t Darmstadt, 64287 Darmstadt, Germany}
\author{A.~Moradabadi}
\affiliation{Institute of Materials Science, Technische Universit\"{a}t Darmstadt, 64287 Darmstadt, Germany}
\affiliation{Institut f\"ur Chemie und Biochemie, Freie Universit\"{a}t Berlin, 14195 Berlin, Germany}
\author{S.~Gorji}
\affiliation{Institute of Nanotechnology (INT), Karlsruhe Institute of Technology (KIT), 76344 Eggenstein-Leopoldshafen, Germany}
\affiliation{Joint Research Laboratory Nanomaterials (KIT and TUD) at Technische Universit\"{a}t Darmstadt, Jovanka-Bontschits-Str. 2, 64287 Darmstadt, Germany}
\author{M.~Yi}
\affiliation{Institute of Materials Science, Technische Universit\"{a}t Darmstadt, 64287 Darmstadt, Germany}
\author{Q.~Gong}
\affiliation{Institute of Materials Science, Technische Universit\"{a}t Darmstadt, 64287 Darmstadt, Germany}
\author{M.~H.~Fawey}
\affiliation{Institute of Nanotechnology (INT), Karlsruhe Institute of Technology (KIT), 76344 Eggenstein-Leopoldshafen, Germany}
\affiliation{Joint Research Laboratory Nanomaterials (KIT and TUD) at Technische Universit\"{a}t Darmstadt, Jovanka-Bontschits-Str. 2, 64287 Darmstadt, Germany}
\author{E.~Hildebrandt}
\affiliation{Institute of Materials Science, Technische Universit\"{a}t Darmstadt, 64287 Darmstadt, Germany}
\author{D.~Wang}
\affiliation{Institute of Nanotechnology (INT), Karlsruhe Institute of Technology (KIT), 76344 Eggenstein-Leopoldshafen, Germany}
\affiliation{Karlsruhe Nano Micro Facility, Karlsruhe Institute of Technology(KIT), 76344 Eggenstein-Leopoldshafen, Germany}
\author{H.~Zhang}
\affiliation{Institute of Materials Science, Technische Universit\"{a}t Darmstadt, 64287 Darmstadt, Germany}
\author{B.~-X.~Xu}
\affiliation{Institute of Materials Science, Technische Universit\"{a}t Darmstadt, 64287 Darmstadt, Germany}
\author{C.~K\"ubel}
\affiliation{Institute of Nanotechnology (INT), Karlsruhe Institute of Technology (KIT), 76344 Eggenstein-Leopoldshafen, Germany}
\affiliation{Karlsruhe Nano Micro Facility, Karlsruhe Institute of Technology(KIT), 76344 Eggenstein-Leopoldshafen, Germany}
\author{L.~Alff}
\email{alff@oxide.tu-darmstadt.de}
\affiliation{Institute of Materials Science, Technische Universit\"{a}t Darmstadt, 64287 Darmstadt, Germany}
\title{\centering On the origin of incoherent magnetic exchange coupling in MnBi/Fe$_{x}$Co$_{1-x}$ bilayer system}

\newpage

\begin{abstract}

In this study we investigate the magnetic exchange coupling behavior in MnBi/FeCo system at the hard/soft interface.  
Exchange spring MnBi/Fe$_{x}$Co$_{1-x}$ ($x=0.65$ and $0.35$) bilayers with various thicknesses of the soft magnetic layer were deposited onto quartz glass substrates in a DC magnetron sputtering unit from alloy targets. 
According to magnetic measurements,  using a Co-rich layer leads to more coherent exchange coupling with optimum soft layer thickness of about 1\,nm. 
In order to take into account the effect of structural factors at the hard/soft interface  
which can deteriorate the exchange coupling for thicker soft magnetic layers, we have combined cross-sectional High Resolution Transmission Electron Microscopy (HR-TEM) analysis with DFT calculations and micromagnetic simulations. DFT calculations predict formation of a polycrystalline FeCo layer with coexisting crystalline (110) and disordered (randomly-oriented) phases which is confirmed by HR-TEM images. 
Moreover, our micromagnetic simulations show how the thickness of the FeCo layer and the interface roughness between the hard and soft magnetic phases both control the effectiveness of exchange coupling in MnBi/FeCo system. Our method can be applied to study other exchange spring systems. 


\end{abstract} 

\maketitle


\section{Introduction}

Exchange spring magnets provide an interesting approach to enable synthesis of rare-earth free permanent magnets with comparable magnetic properties to commerialized rare-earth based magnets. The price instability of rare-earth resources has made the applications of Nd and Sm based magnets economically critical. There is an urgent need to develop permanent magnets with reduced rare-earth contents or to explore rare-earth free permanent magnetic materials where Mn-based intermetallics are considered to be potential candidates \cite{Poudyal:13, Kramer:12,DLi:16}. In order to be qualified as replacements for rare-earth magnets, it is required that new candidates have high magnetic anisotropy, high energy product, and high temperature stability \cite{Gutfleisch:11}.
The low temperature phase (LTP) of MnBi is one of the materials with a particularly high intrinsic magnetic anisotropy in the order of $10^7$\,erg/cm$^3$ as well as a large coercivity (about 1.6\,T), which rather uniquely shows a positive temperature coefficient \cite{Sabet:17,Park:14,Suzuki:15}.
Moreover, the relatively high Curie-temperature of 630\,K also makes MnBi an interesting candidate for high temperature applications \cite{Guillaud:51a}.
However, in spite of such extraordinary magnetic properties, the main drawback of MnBi for permanent magnet application
is its comparably low saturation magnetization of 710\,emu/cm$^3$ (0.71\,MA/m) limiting the maximum achievable energy product \cite{Park:14}.

\begin{figure}[!th]
	\centering
	\includegraphics[width=0.49\textwidth]{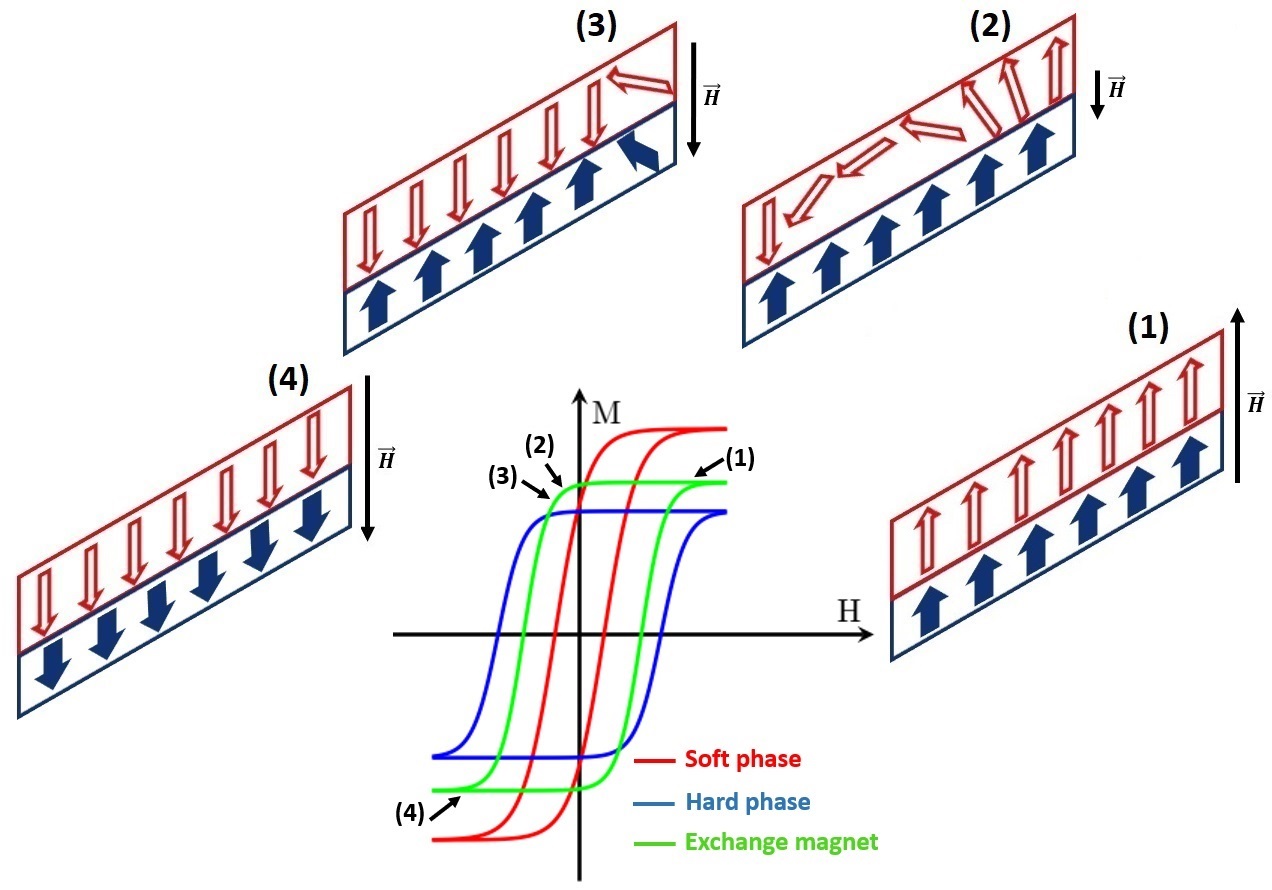}
	\caption{ Schematic of typical hysteresis loops for a hard magnetic phase (e.g. MnBi, blue solid line), a soft magnetic phase (e.g. FeCo, red solid line), and the exchange spring composite magnet (e.g. MnBi/FeCo, green solid line) resulting from hard/soft exchange coupling. The four insets show a one-dimensional configuration of magnetic moments at the hard/soft magnetic interface under varying external magnetic field ($H$) assuming an out-of-plane easy axis direction. Red open arrows represent the magnetic moments in the soft phase and blue filled arrows represent the magnetic moments in the hard phase. The length of the arrows represents the magnetization and the width of the arrows represents the coercivity.}
	\label{EX}
\end{figure}

As suggested back in 1991 by Kneller and Hawig, one way to overcome this barrier and further improve the energy product is through the synthesis of exchange spring magnets with coupled hard/soft magnetic phases (see schematics in Fig. \ref{EX}) \cite{Kneller:91,Fullerton:98,Fullerton:99,Leineweber:97,Coey:93,Skomski:94,Lewis:94}.
Such composite magnets, $e.g.$ coupled bilayers of MnBi in combination with FeCo as the soft phase, will possess a much higher saturation magnetization and thus an increased overall energy product.

For exchange spring heterostructures with MnBi as hard magnetic phase, to our knowledge there have been only a few recent studies investigating the synthesis and magnetic properties of the resulting bilayers \cite{Li:14,Gao:16,Yan:16,Sabet:17a}.
Although theoretical calculations have proven the concept of exchange spring magnets to increase the overall magnetic properties, according to the few available studies the coupling between the MnBi and Fe$_{x}$Co$_{1-x}$ layers is incoherent for magnetic layers thicker than $\sim$ 4\,nm \cite{Li:14,Gao:16,Yan:16,Sabet:17a}. This is also evident from the small shoulder observed around zero field on the hysteresis loops measured in all above mentioned studies where the two layers do not behave as a single magnetic phase. It is important to understand the interfacial effects responsible for an incoherent interlayer exchange coupling in order to make further advances to effectiveness of exchange spring magnets.

Based on the model suggested by Kneller \cite{Kneller:91}, there is a critical thickness (volume) of soft magnetic phase which is limited by the domain wall width (or exchange length) of the hard magnetic phase \cite{Fullerton:99, Leineweber:97}.
For thicker soft magnetic layers, the coupling between hard and soft magnets begins to deteriorate and hence the layers will switch independently during the magnetic reversal process.
However, if the thickness of the soft magnetic layer is less than twice of the domain wall width of the hard magnetic phase, the bilayer is expected to behave as a single hard phase with increased magnetization in which both soft and hard phases switch coherently during magnetic reversal under opposing field ($H<0$).
Experimentally, even for sufficiently thin soft magnetic layers incomplete exchange coupling is reported indicating that other factors are involved.

Beside the thickness of the soft and hard magnetic layers, structural factors such as degree of crystallinity and growth orientation are expected to affect the strength of exchange coupling. For instance, the lattice mismatch at the interface can act either in favor or against magnetic coupling. The hard/soft interface roughness resulting from the growth quality of the layers can influence the coupling between the layers as well. In addition, the effect of composition of the Fe$_{x}$Co$_{1-x}$ soft magnetic layer has been considered as a controlling factor affecting the interlayer exchange coupling. Based on their calculations, Gao {\it et al.} have also argued that the formation of a Co-rich Fe$_{x}$Co$_{1-x}$ layer at the interface with MnBi is beneficial for exchange coupling where according to their experimental data the strongest coupling occurs in MnBi/Co bilayers with an optimum Co thickness of $\sim$ 3\,nm \cite{Gao:16}.

In this work we combined theoretical and experimental methods to study the exchange coupling behavior in the MnBi/FeCo bilayer system, focusing on the structural factors including the effect of degree of crystallinity, interface roughness and composition of the soft magnetic phase to identify which of these factors control the strength of exchange spring effect in this system.

\section{Experimental procedure}

Exchange coupled bilayers of MnBi/FeCo have been deposited onto quartz glass substrates in a DC magnetron sputtering system with a base pressure of $\sim$ 4.0$\times10^{-6}$\,Pa and were capped with 4\,nm thick aluminium layer to protect them against oxidation. First, a MnBi layer with a typical thickness of 40\,nm was deposited from a Mn-Bi alloy target with a composition of Mn$_{55}$Bi$_{45}$ (at.\%) at room temperature. The optimized growth parameters in our setup were 0.7\,Pa Ar gas pressure at 20\,W sputtering power with a substrate to source distance of 15\,cm leading to a growth rate of 0.04\,nm/s. The MnBi film was subsequently annealed for 1\,hr (dwelling time) {\em in situ} under vacuum ($\sim 1.0\times10^{-5}$\,Pa) at the annealing temperature of $T_{\text{ann}}= 365$\,\degree C. The temperature was ramped up and down with a rate of 20\,\degree C/min and 10\,\degree C/min, respectively. After cooling to 100\degree C, the soft magnetic layers using either a Fe-rich or a Co-rich, Fe$_{x}$Co$_{1-x}$ ($x=0.35$ or $0.65$) alloy target with various thicknesses of 1\,nm-3\,nm were deposited on top of the MnBi layer. The growth parameters for FeCo deposition were 2.5\,Pa Ar gas pressure, 80\,W sputtering power and 8\,cm substrate to source distance leading to a deposition rate of 0.008\,nm/s. Then the substrate temperature was increased to 120\degree C for deposition of an aluminium capping layer. The capping layer was deposited under 3\,Pa Ar gas pressure at 20\,W sputtering power with 15\,cm substrate to source distance at a rate of $\sim$0.008\,nm/s. The phase composition and degree of texture for the MnBi layer were determined by X-ray diffraction with Cu-$K_{\alpha}$ radiation using a Rigaku SmartLab thin film diffractometer. The film thickness was determined by a Bruker Dektak-XT stylus surface profiling system. The magnetic properties were measured by a SQUID magnetometer (MPMS, QuantumDesign).
For cross-sectional High Resolution Transmission Electron Microscopy (HR-TEM) investigations, TEM lamella was prepared by Focused Ion Beam (FIB) using a FEI Strata 400S equipped with an OmniProbe 200 micromanipulator for $in$-$situ$ lift-out. TEM sample preparation was initially performed at 30\,kV with an ion beam current of 16\,nA, followed by cleaning with a 6.5\,nA ion beam current. The final thinning step of the area of interest at the interface was performed at a low voltage-low current regime starting from 8\, to 2\,kV with ion beam current ranging from 56\,pA to 3\,pA. An aberration (image) corrected FEI Titan 80-300 operating at 300\,kV acceleration voltage and equipped with a US1000 slow-scan CCD camera (Gatan Inc.), a high-angle annular dark-field (HAADF) detector (Fischione), and an S-UTW EDX detector (EDAX Inc.) were used to evaluate the crystallinity, interface quality and composition.

\section{Theoretical procedure}
\subsection{Density functional theory}
Density functional theory (DFT) calculations were performed using the projected augmented wave method as implemented in the
\texttt{VASP} code~\cite{vasp}. 
The exchange correlation functionals were parameterized using the generalized gradient approximation (GGA) as in Ref.~\cite{PBE}. The effect of the GGA+U approximation, which is important to understand the bulk anisotropy of MnBi \cite{pbe+u} was investigated and no significant influence on the interface properties was observed. 
Two MnBi/FeCo models were considered in our calculations, namely MnBi(001)/FeCo(110) and MnBi(001)/FeCo(111), each in both crystalline and amorphous states.
The amorphous structures were generated using {\it ab initio}-based molecular dynamics (AIMD) calculations implemented in \texttt{VASP} after 10\,ps run for the FeCo(111) and 2\,ps run for the FeCo (110) surfaces, respectively, both at 500$^{\circ}$C.
All models were constructed as symmetric and non-stoichiometric slabs for their interface formation energies would be comparable.
Moreover, Fe$_x$Co$_{(1-x)}$ layers with two different stoichiometries, {\it i.e.} Fe$_3$Co$_5$ and Fe$_5$Co$_3$, were considered in order to study the influence of chemical composition on the interfacial properties. 
The thickness of the MnBi layer in MnBi(001)/FeCo(110) and MnBi(001)/FeCo(111) interfaces was 10 \AA ~and 15 \AA, respectively.
During the calculations, the first two layers of MnBi were kept fixed while the rest of the MnBi and the whole FeCo layers were fully relaxed. At least 14 \AA~vacuum was considered when constructing the supercells using slab models to minimize the interaction between periodic images.
Interface formation energy $\gamma_{\rm int}$, which is a measure of the stability of the corresponding interface, was calculated using the following equation
\begin{equation} \label{eq1}
	\gamma_{\rm int}=\frac{1}{2S}\bigg[E_{\rm int}-n_1E^{\rm MnBi}_{\rm bulk}-n_2E^{\rm FeCo}_{\rm bulk}+\sum_i\mu_i\bigg]
\end{equation}
where $S$ and $E_{\rm int}$ are the area and the total energy of the whole interface, $E^{\rm MnBi}_{\rm bulk}$ and $E^{\rm FeCo}_{\rm bulk}$ are the total energies per formula unit of MnBi and FeCo, and $n_1$ and $n_2$ are the number of bulk units of MnBi and FeCo in the models, respectively. $\mu_i$ is the chemical potential of any  missing atoms summation of which maintains the stoichiometry. The chemical potentials were considered as the total energies per atom in metallic bulks. 
The interface exchange coupling energy $J^\text{int}$ was obtained using the following relation 
\begin{equation} \label{eq2}
	J^\text{int}=(E^\text{APMA}-E^\text{PMA})/S
\end{equation}
where $E^\text{APMA}$ and $E^\text{PMA}$ are the  
DFT total energies for antiparallel magnetization alignment (APMA) and parallel magnetization alignment (PMA), respectively.


Based on the optimized lattice parameters using DFT calculations, 
the lattice mismatch between the MnBi(001) and FeCo(110) is 4.8\% with a 10.5$^{\circ}$ angular misfit while the lattice mismatch between MnBi(001) and FeCo(111) is 7.1\% with zero angular misfit. 
Our MnBi(001)/FeCo(111) model consisted of a 2$\times$2 supercell of MnBi(001) and a 1$\times$1 supercell of FeCo(111) with 96 atoms in total. 
In our MnBi(001)/FeCo(110) model, we used a 6$\times$6 supercell of MnBi(001) together with a 5$\times$5 supercell of FeCo(110) with 752 atoms in total to achieve the 4.8\% lattice misfit. 
For the calculation of MnBi(001)/FeCo(111) interface, 3$\times$3$\times$1 $k$-point mesh was used and the calculation of MnBi(001)/FeCo(110) was performed with gamma-point. The energy cutoff for all the calculations was 360 eV. The convergence tests of the energy cutoff and $k$-point mesh with respect to the magnetic moments of the elements in their bulk states and energy per atom were conducted.  

\subsection{Micromagnetic simulation}

Using the results from DFT calculations as input, micromagnetic simulations were performed within a simplified model to investigate the mechanism of exchange coupling in MnBi/FeCo magnets by using the 3D NIST OOMMF (Object Oriented Micromagnetic Framework) code \cite{oommf}. In our model, the thicknesses of the hard MnBi and soft FeCo layers (initially) were set as 40\,nm and 2\,nm, respectively. The lateral size is chosen as $8\times 8$ nm$^2$ and an in-plane periodic boundary condition was applied. The model was discretized by 0.4\,nm$\times$0.4\,nm$\times$0.1\,nm cuboid cells. Magnetic reversal curves were calculated by setting the initial magnetization along a positive $z$ axis and changing the external magnetic field along $z$ axis from 2.5\,T to $-2.5$\,T. 

The exchange stiffness $A^\text{int}$ which characterizes the exchange coupling between MnBi and FeCo layers through the interface was used. According to the method of calculating exchange energy in OOMMF code \cite{oommf}, interface exchange stiffness $A^\text{int}$ was estimated by the following expression 
\begin{equation}
A^{\rm int} \cong I^\text{vol}\Delta z^2/2
\end{equation}
in which $\Delta z=0.1$ nm is the cell size in the $z$ direction and $I^\text{vol}$ is the equivalent volumetric energy density calculated by $J^\text{int}$ divided by the average interface distance measured from the crystal structures after relaxation from our DFT calculations.
The resulting $\gamma^{\rm int}$, $J^{\rm int}$ and $I^{\rm int}$ values are shown in Table \ref{t1}.
The bulk parameters for exchange stiffness $A$ and uniaxial anisotropy constant $K$ were set as: $A^\text{FeCo}=10$ pJ/m, $A^\text{MnBi}=8$ pJ/m, $K^\text{FeCo}=0$ MJ/m$^3$, and $K^\text{MnBi}=1.86$ MJ/m$^3$~\cite{Rana:16,Sabet:17}. 
The saturation magnetizations, $M_\text{s}^\text{FeCo}$ and $M_\text{s}^\text{MnBi}$, were also obtained from the DFT results. 

\section{Results and discussion}

\begin{figure}[!th]
	\centering
	\includegraphics[width=0.62\textwidth]{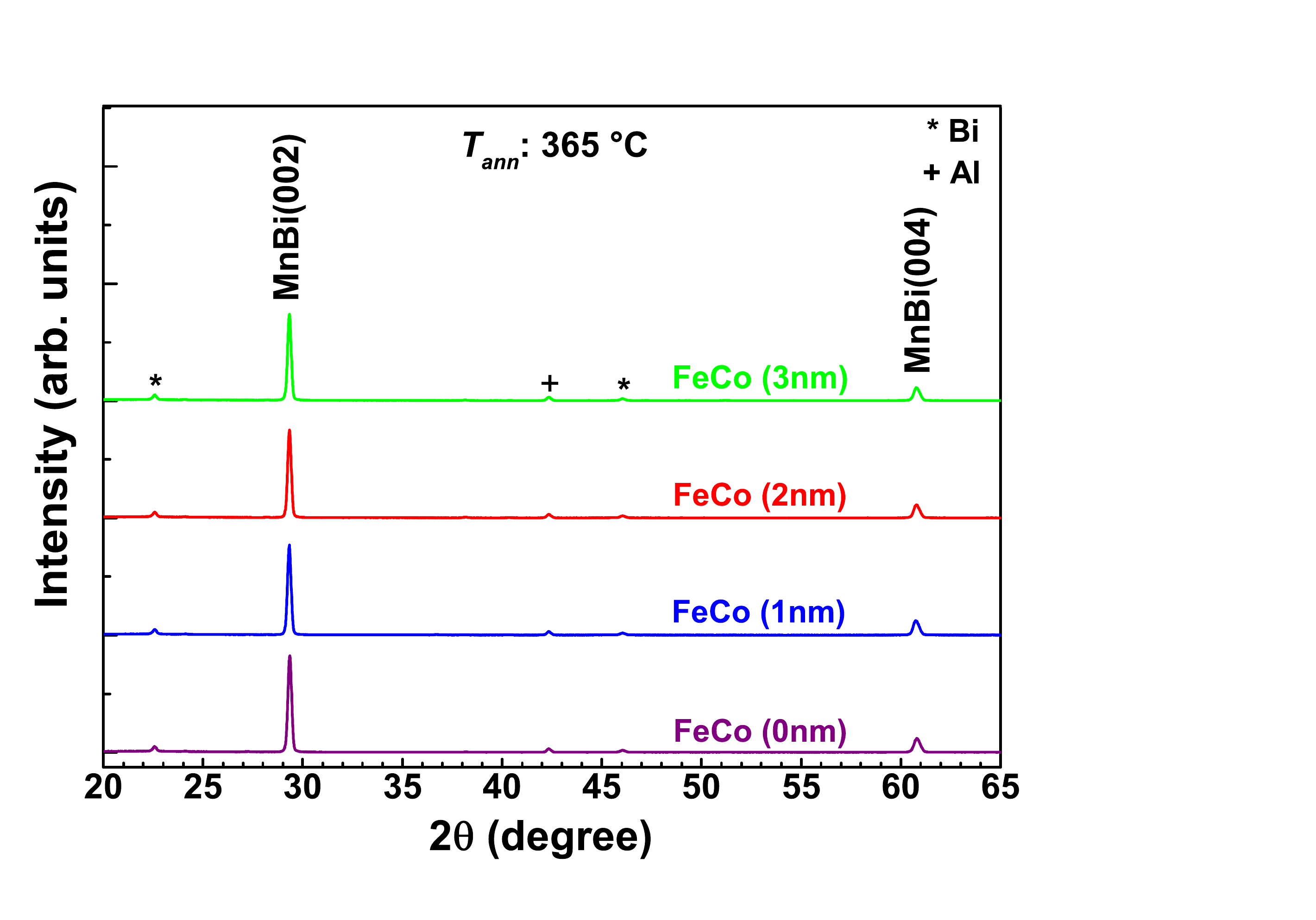}
	\caption{ The XRD patterns from exchange spring bilayers of Mn$_{55}$Bi$_{45}$/Fe$_{35}$Co$_{65}$ (at.\%) with different thickness of FeCo soft magnetic layer between 1\,nm-3\,nm. The LTP-MnBi thin films were annealed at $T_{\rm ann}$=365\,\degree C followed by deposition of FeCo layer at a substrate temperature of $T_{\rm sub}$=100\,\degree C. The spectra have vertical offset for clarity. The peaks originating from residual bismuth in the films or Al capping layer are labelled with (*) and (+), respectively.}
	\label{XRD-MB-FC}
\end{figure}

The XRD patterns collected from different Mn$_{55}$Bi$_{45}$/Fe$_{35}$Co$_{65}$ (at.\%) exchange spring bilayers with various thicknesses of Co-rich soft magnetic FeCo layer are shown in Fig.~\ref{XRD-MB-FC}. The peak indexing shows hexagonal MnBi (002) and (004) peaks in agreement with space group of $P$63/$mmc$ along with some small traces of residual bismuth resulting from annealing of the MnBi films at $T_{\rm ann}=360$\,\degree C. Fig.~\ref{XRD-MB-FC} clearly demonstrates the formation of LTP MnBi with strong $c$-axis texture. As expected, because of the very low thicknesses, no peaks are observed for the FeCo layer. Comparing the intensities of the MnBi (002) and (004) peaks in bilayers samples to that of the single layer MnBi thin film, all the XRD patterns show similar peak intensities implying that the MnBi hard magnetic layer in all the bilayer samples had the same high crystalline quality.
\begin{figure}[!th]
	\centering
	\includegraphics[width=0.55\textwidth]{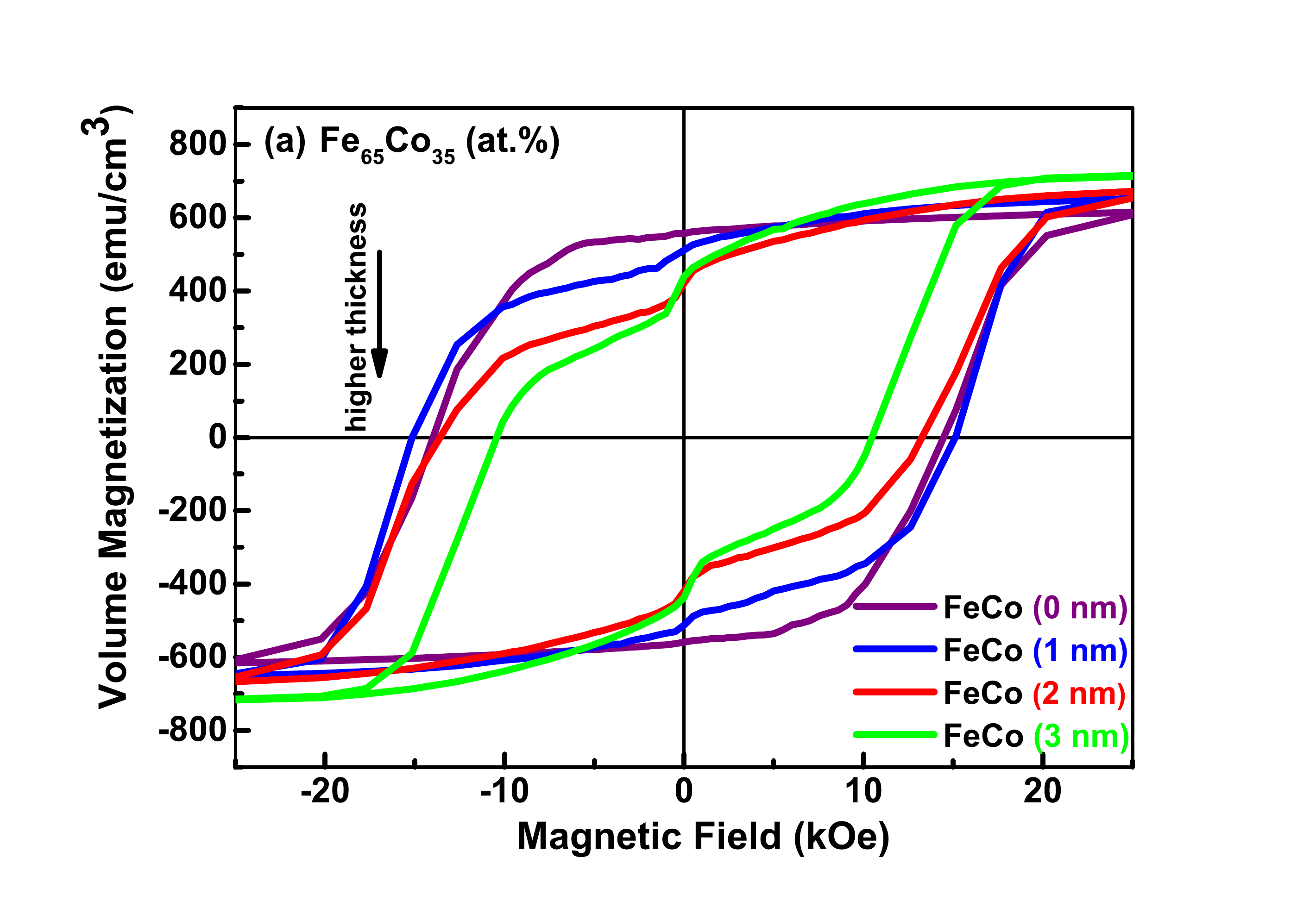}
	\includegraphics[width=0.55\textwidth]{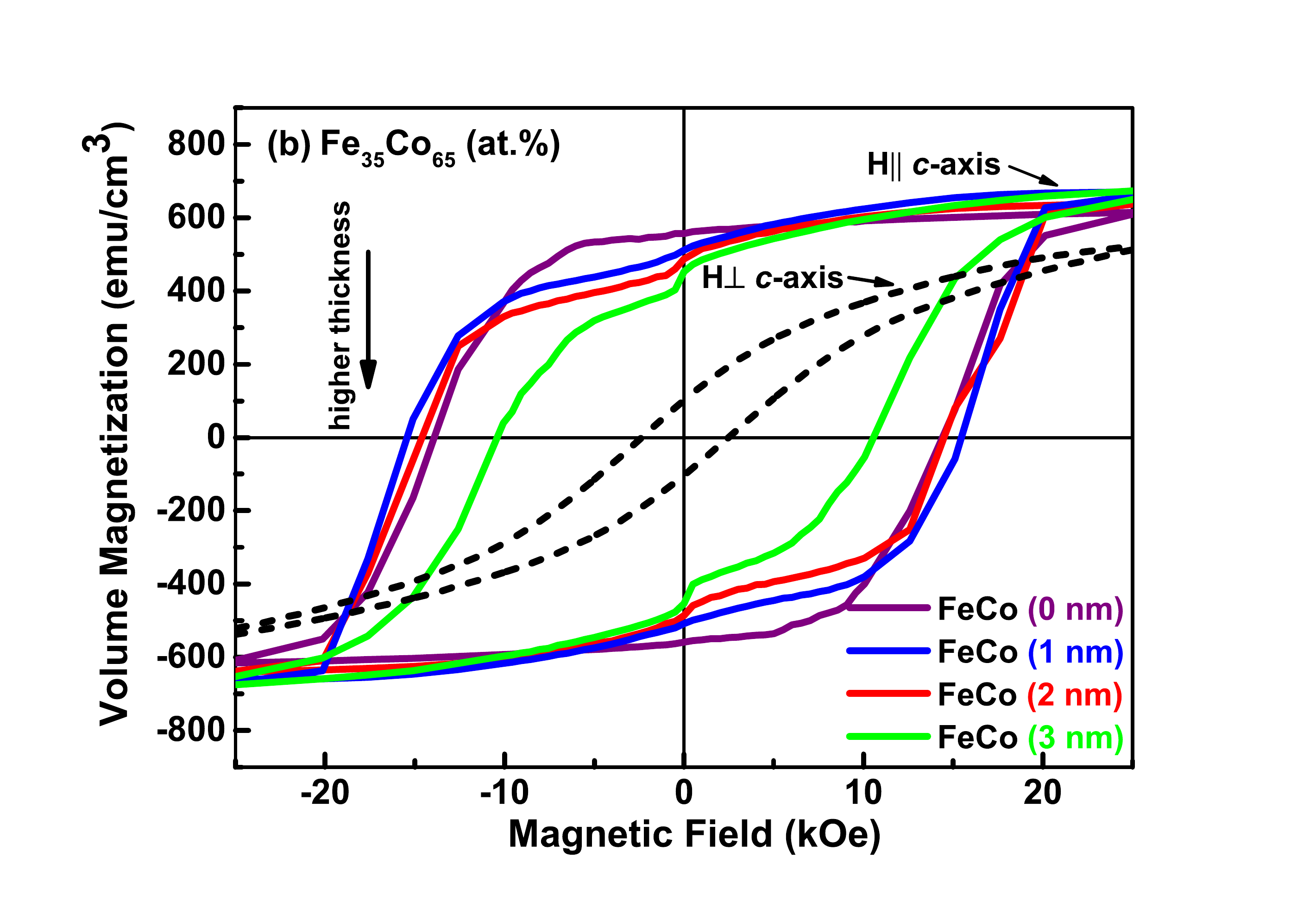}
	\caption{Out-of-plane magnetization data for MnBi/FeCo bilayers with different FeCo thicknesses from 0\,nm to 3\,nm measured at 300\,K, (a) with a Fe-rich and (b) with a Co-rich soft magnetic FeCo layer. The dashed line in (b) shows the in-plane magnetization for a single MnBi layer. }
	\label{MB-FC-M-H}
\end{figure}

Room-temperature out-of-plane hysteresis loops for MnBi/FeCo bilayer samples with various thicknesses and two compositions of the soft magnetic layer are shown in Fig.~\ref{MB-FC-M-H}-a and -b. For comparison, the out-of-plane hysteresis loop for a single layer MnBi thin film sample is also included in the same graph. As expected, by addition of 1\,nm, 2\,nm and 3\,nm FeCo layers for both Fe-rich and Co-rich compositions, the saturation magnetization of exchange spring bilayer increased.
According to the graphs in Fig.~\ref{MB-FC-M-H}, the addition of Fe-rich soft magnetic FeCo layers improved the saturation magnetization more than the addition of Co-rich FeCo layers, since the Fe$_{65}$Co$_{35}$ (at.\%) phase has a $\sim$ 20\% larger saturation magnetization than the Fe$_{35}$Co$_{65}$ (at.\%) phase \cite{Kuhrt:93}. The total magnetization in a bilayer is  given by the volume average of magnetization in the hard and soft magnetic layer \cite{Fullerton:99, Skomski:94}. The deposition of a 1\,nm and 2\,nm thick soft magnetic layer on top of MnBi retains the coercivity of the LTP-MnBi layer (about 15\,kOe, even with a slight increase), while regardless of the composition of the soft magnetic layer the addition of 3\,nm FeCo decreases the coercivity down to 12\,kOe.

The exchange coupling effect between the hard and soft magnetic layers can be considered complete when the bilayer sample shows a magnetically single phase behavior. 
The small shoulder which was observed during the demagnetization process around zero field in the measured out-of-plane hysteresis curves of the double layers indicates that the exchange coupling between the layers is incoherent. As a descriptor to quantify the change in the degree of exchange coupling, the slope ($\frac{\Delta M}{\Delta H}$) of the hysteresis loop around zero-field crossing has been evaluated. 
This will be explained in the following with more details using the micromagnetic simulations. As this slope increases with growing a thicker soft magnetic layer, it implies that the bilayers behave more as two separate magnetic layers instead of one single magnetic phase. 
Although such decrease in degree of coupling is predicted with increasing thickness of the soft magnetic layer, it is also expected that the critical soft layer thickness, above which the exchange coupling begins to deteriorate, is roughly twice the domain wall width of the hard magnetic layer ($2\times \delta_{h}\simeq 2\times\pi\sqrt{\frac{A_{h}}{K_{h}}})$ \cite{Kneller:91, Fullerton:99,Leineweber:97} in which $\delta_{h}$ is domain wall width, $A_h$ is exchange stiffness constant and $K_h$ is magnetocrystalline anisotropy for the hard magnetic phase. 
For a MnBi-FeCo bilayer with $A_h$ and $K_h$ equal to $\sim 1.0\times10^{-6}$\,erg/cm \cite{Rana:16} and $\sim 1.86\times10^{7}$\,erg/cm$^{3}$ \cite{Sabet:17} respectively, the critical thickness is predicted to be as high as $\sim$15\,nm. It becomes obvious that one needs to take into account a more detailed interface description to explain the experimental observations.

The observed incoherent coupling in the MnBi/FeCo exchange spring system can be attributed to different structural factors including: (i) a non-epitaxial hard magnetic layer which most likely results in subsequent growth of a polycrystalline or disordered soft magnetic layers on top, (ii) high interface roughness which could also be a side effect of non-epitaxial growth of the layers,  caused by lattice mismatch at the interface, and (iii) composition gradients in the FeCo layer in the vicinity of the interface. In addition, according to Fig. \ref{MB-FC-M-H}, there exists a finite in-plane component of total magnetization in the MnBi hard magnetic layer. The in-plane components of the magnetization when incompletely or not coupled lead to a kink at the coercive field of the soft magnetic layer($H$ close to zero).

\begin{figure}[!ht]
	\centering
	\includegraphics[width=0.49\textwidth]{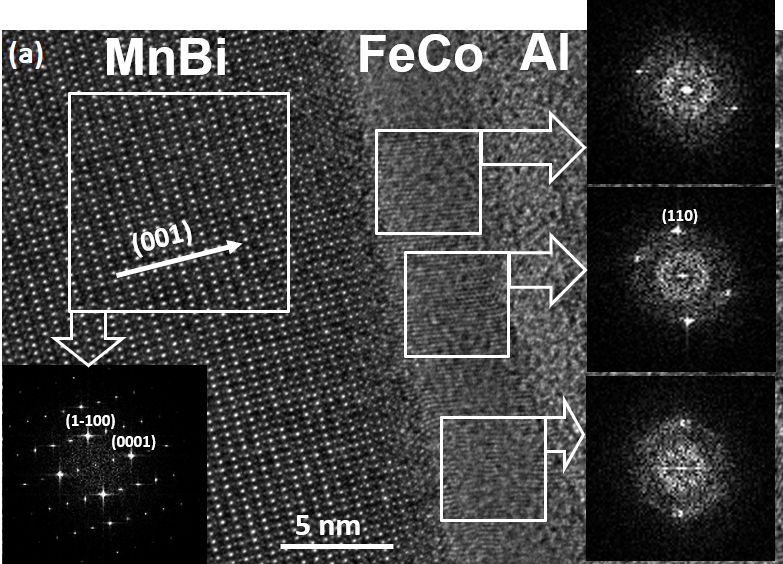}
	\includegraphics[width=0.30\textwidth]{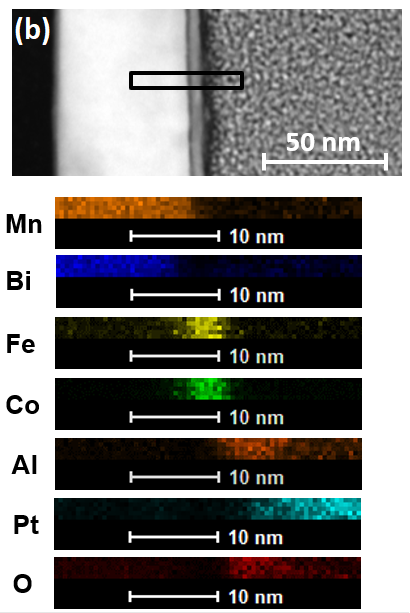}
	\caption{(a) Cross-sectional High Resolution Transmission Electron Microscopy (HR-TEM) image from a MnBi/FeCo bilayer sample( $c$-axis textured MnBi hard magnetic layer with a thickness of $\sim$ 50\,nm and polycrystalline Co-rich FeCo soft magnetic layer with a thickness of $\sim$ 5\,nm), (b) STEM image from cross section of the layers along with EDX elemental map from Mn, Bi, Fe, Co, Al, Pt and O across the layers.}
	\label{MB-FC-TEM}
\end{figure}

To examine the interface between MnBi and FeCo, cross-sectional HR-TEM and STEM investigations have been performed on a MnBi/FeCo bilayer sample. The degree of crystallinity was evaluated for the layers by capturing HR-TEM images from cross-section of the layers. Moreover, the distribution of different elements in each layer was examined in STEM mode.
Fig.~\ref{MB-FC-TEM}-a shows a cross-sectional HR-TEM image of the layers along with Fast Fourier Transform (FFTs) collected from each layer in a bilayer sample with a Co-rich soft layer. The HR-TEM image and the sharp diffraction spots in FFTs collected from the MnBi layer confirm the high crystallinity with out-of-plane orientation. The Co-rich FeCo layer, in contrary, shows polycrystalline structure. 

Three different surface areas have been analyzed in the FeCo layer with various crystallinity. Only the examined area in the middle FFT shows high crystallinity and the two other investigated regions are disordered. 
The reflections in the middle FFT pattern (FFT-A2 in Fig.~\ref{MB-FC-TEM}-a) of the crystalline region in FeCo layer can be indexed as (110) lattice plane.
As it can be seen in the cross-section HR-TEM image of the MnBi/FeCo bilayer, a few atomic layers of FeCo layer grown on MnBi are highly disordered. This was expected since FeCo and the (001) textured MnBi layer have different crystal structures, i.e.  hexagonal structure in MnBi and bcc structure in FeCo, which results in the growth of polycrystalline FeCo layer because of the induced lattice misfit. The imperfection of crystallinity and the existence of grain boundaries in the FeCo layer also leads to the formation of a rough interface.

To check the elemental distribution in the bilayer sample, an EDX mapping was performed on the enclosed area in Fig.~\ref{MB-FC-TEM}b. The result of the EDX mapping is consistent with the phases present in each layer. Close to the interface between the two layers the Bi concentration starts to decrease earlier than the Mn concentration. 
According to the quantitative EDX analysis from this specific area on the cross-section of the bilayer sample, the MnBi layer shows a stoichiometry of Mn:Bi$\sim$ 1.4 which is slightly higher than the starting stoichiometry of 1.2 in the alloy sputtering target. This value corresponds to a final stoichiometry of Mn$_{58}$Bi$_{42}$ (at.\%) which is slightly richer in Mn. The measured stoichiometry for the FeCo layer shows a Co:Fe ratio of $\sim$ 1.84 which is consistent with the starting stoichiometry of 1.86 in the alloy sputtering target. This confirms a fairly precise stoichiometry transfer from MnBi and FeCo alloy sputtering targets during film deposition.

\begin{figure}[!th]
  	\centering
  	(a)\\
  	\includegraphics[width=0.45\textwidth]{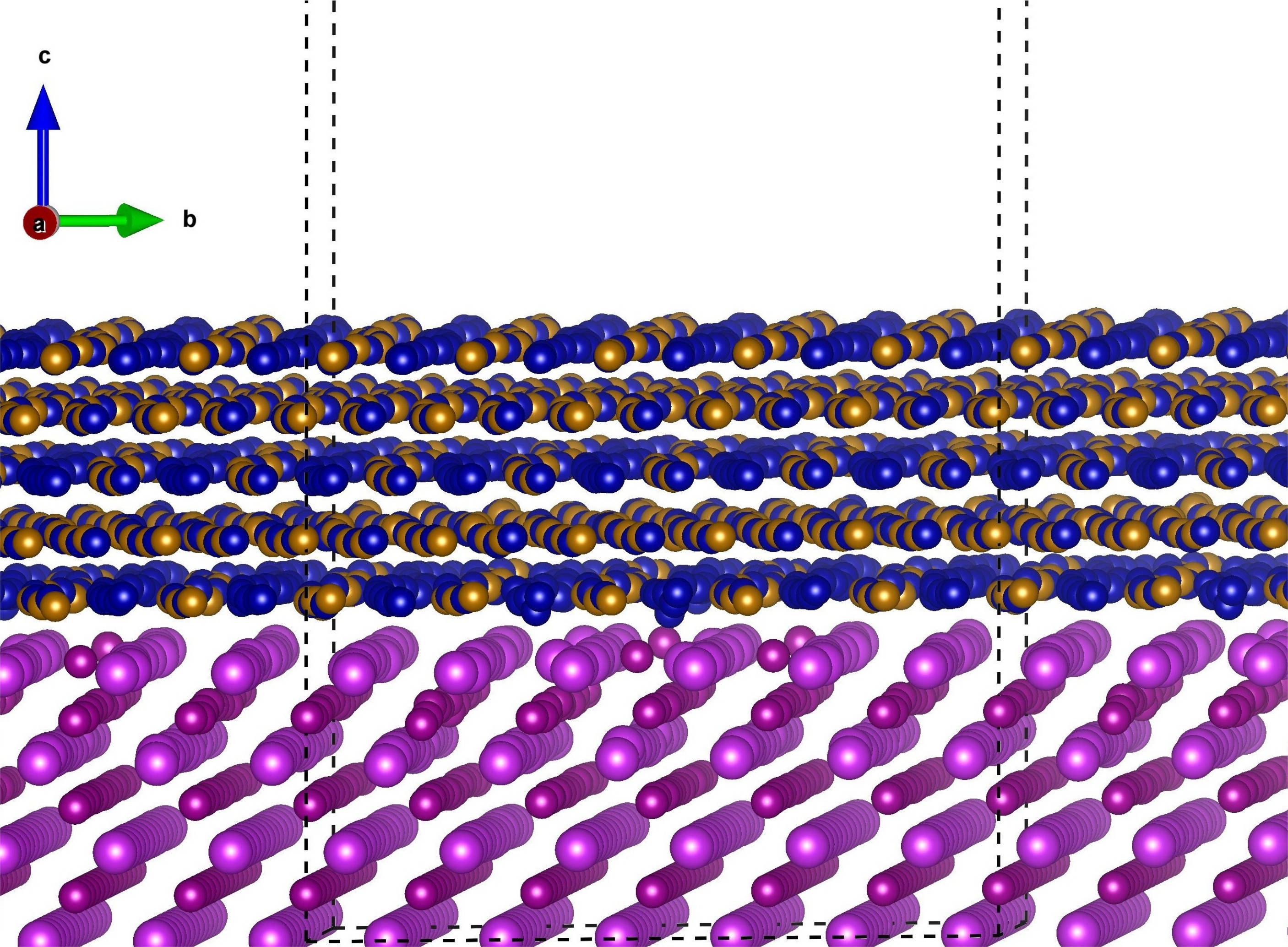}\\
  	\vspace*{0.5cm}
  	(b)\\
  	\includegraphics[width=0.45\textwidth]{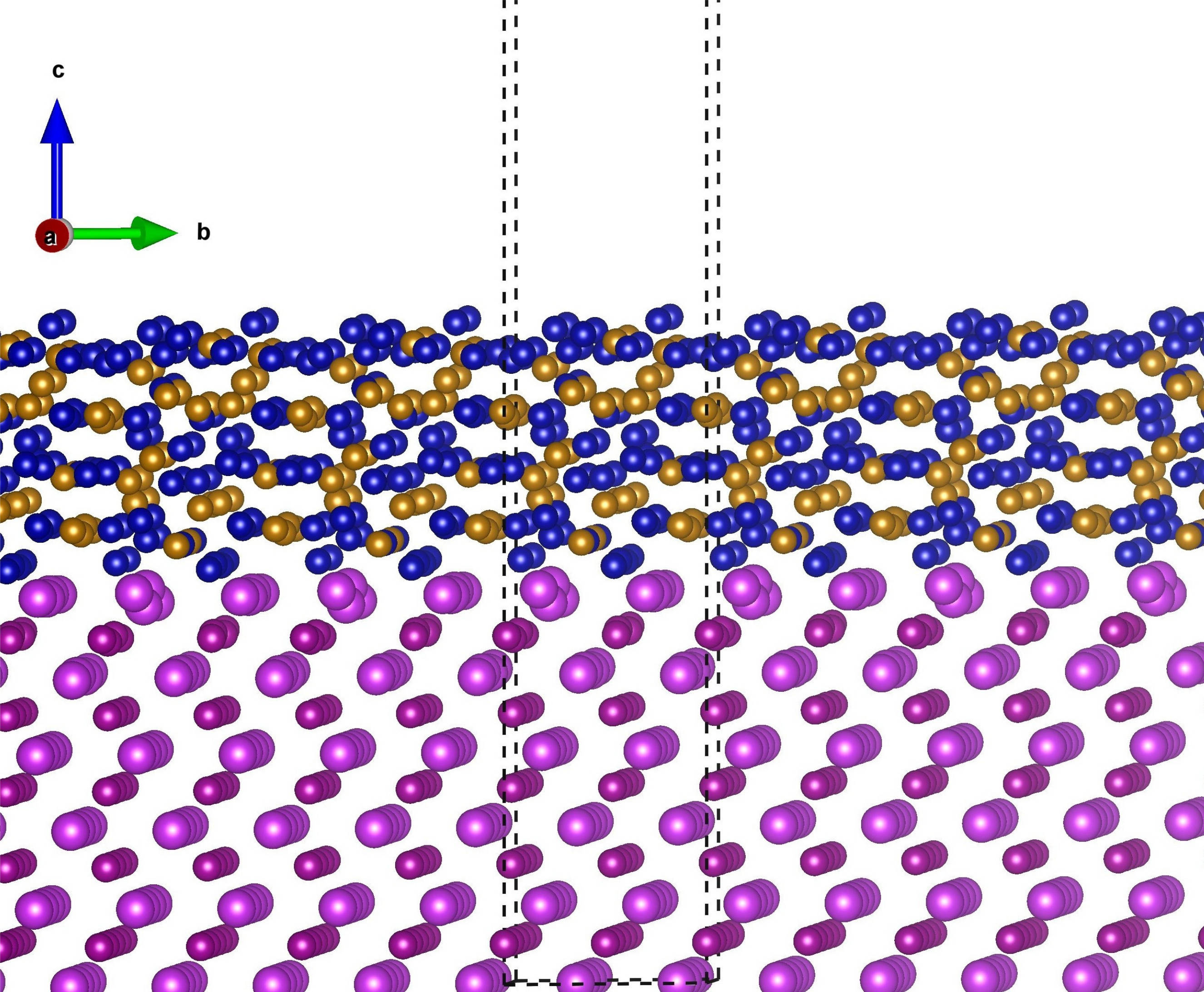}\\
  	\caption{Schematic of atomic structures of relaxed (a) MnBi(001)/crystalline FeCo(110) and (b) MnBi(001)/amorphous FeCo(111) interfaces demonstrated using VESTA~\cite{vesta}. Mn, Bi, Fe and Co are shown with small dark violet, large light violet, gold and blue colors, respectively. The  1$\times$1$\times$1 unit cell of each orientation is indicated with dashed line.}
  	\label{str}
  \end{figure}

In order to shed light on the possible mechanism which affects the performance of the MnBi/FeCo exchange spring magnets, density functional theory (DFT) calculations and micromagnetic simulations were carried out, with a focus on the interface properties.
Fig.~\ref{str} shows the atomic structures of the most favorable configurations after the atomic relaxation
for MnBi(001)/crystalline FeCo(110) (after atomic relaxation MnBi(001)/disordered FeCo(110)) and  MnBi(001)/amorphous FeCo(111) interfaces. 

  \begin{table*}[!tbh]
  	\centering
  	\caption{{Calculated values of interface formation energy $\gamma^\text{int}$, interface  exchange coupling energy $J^\text{int}$ , exchange constant $A^\text{int}$ and lattice misfit (linear and angular) obtained from DFT calculations. 
  			Disordered structure is reconstructed while amorphous one is completely irregular.}}
  	\begin{tabular}{ccccccc}
  		\hline\hline\\
  		Composition & Lattice misfit & Orientation & \makecell {Final phase \\ after relaxation } & \makecell {$\gamma^{\rm int}$ \\ (eV/\AA$^2$)} & \makecell{$J^\text{int}$ \\ (J/m$^2$)} &   \makecell{$A^\text{int}$ \\ (pJ/m)} \\ \hline\\
  		Fe$_3$Co$_5$ & \multirow{2}{*}{7.1\%, 0$^{\circ}$} &(111) & crystalline & 0.137 & 0.112 & 4.6 \\
  		Fe$_3$Co$_5$ &  & (111) & amorphous & 0.130 & 0.093 & 3.3 \\
  		\cline{1-7}\\
  		Fe$_3$Co$_5$ & \multirow{2}{*}{4.8\%, 10.5$^{\circ}$} &(110) & crystalline & 0.129 & 0.195 & 4.5 \\
  		Fe$_3$Co$_5$ &  & (110) & disordered & 0.127 & 0.073 & 1.7 \\
  		\hline\hline\\
  	\end{tabular}
  	\label{t1}
  \end{table*} 

First, it is found that in the MnBi/FeCo bilayer system, the most favourable atomic configuration at the interface and 0 K temperature forms with  Bi-termination MnBi and Co-termination FeCo which is obtained with a symmetric non-stoichiometric model. These findings are in agreement with the cohesive energies of these elements \cite{kittel}. However, our DFT calculations show that the interchange of one Mn atom from MnBi and one Fe atom from FeCo requires only 0.3 eV energy which is a rather low energy barrier and can show the possibility of Mn migration across the interface. The presence of Mn atoms close to the interface is also observed in the EDX elemental map (see Fig. \ref{MB-FC-TEM}-b).

Second, based on the results shown in Tab.~\ref{t1}, it is demonstrated that 
 the interface formation energies of the MnBi(001)/crystalline Fe$_3$Co$_5$(110) and MnBi(001)/disordered Fe$_3$Co$_5$(110) interfaces are almost the same and the formation energy of the former is lower than MnBi(001)/crystalline Fe$_3$Co$_5$(111) case. This can be related to the fact that MnBi(001)/ Fe$_3$Co$_5$(110) interface has lower lattice mismatch compared to MnBi(001)/ Fe$_3$Co$_5$(111) case.
It should also be noted that during the atomic relaxation of the MnBi(001)/ crystalline FeCo(110) interface, the FeCo layer undergoes an atomic reconstruction with a peculiar spiral fashion which has the minimum energy configuration. Moreover, the final ground state structure of the MnBi(001)/amorphous Fe$_3$Co$_5$(110) interface obtained from AIMD calculation is not stable and transforms into a more ordered (almost crystalline) state which is opposite to the case of the MnBi(001)/amorphous Fe$_3$Co$_5$(111) interface. 


Third, the similar and low interface formation energies of the  MnBi(001)/amorphous Fe$_3$Co$_5$(111),  MnBi(001)/crystalline Fe$_3$Co$_5$(110) and MnBi(001)/disordered Fe$_3$Co$_5$(110) suggest the possible coexistence of the crystalline(110) and disordered  structures at the interface region on FeCo side. Interestingly, both crytsalline FeCo(110) and disordered (randomly oriented) regions have been observed in our  cross-sectional HR-TEM image (Fig. \ref{MB-FC-TEM}-a) at the FeCo side which is in agreement with the result of DFT calculations.

 \begin{figure*}[!th]
 	\centering
 	\includegraphics[width=1.0\textwidth]{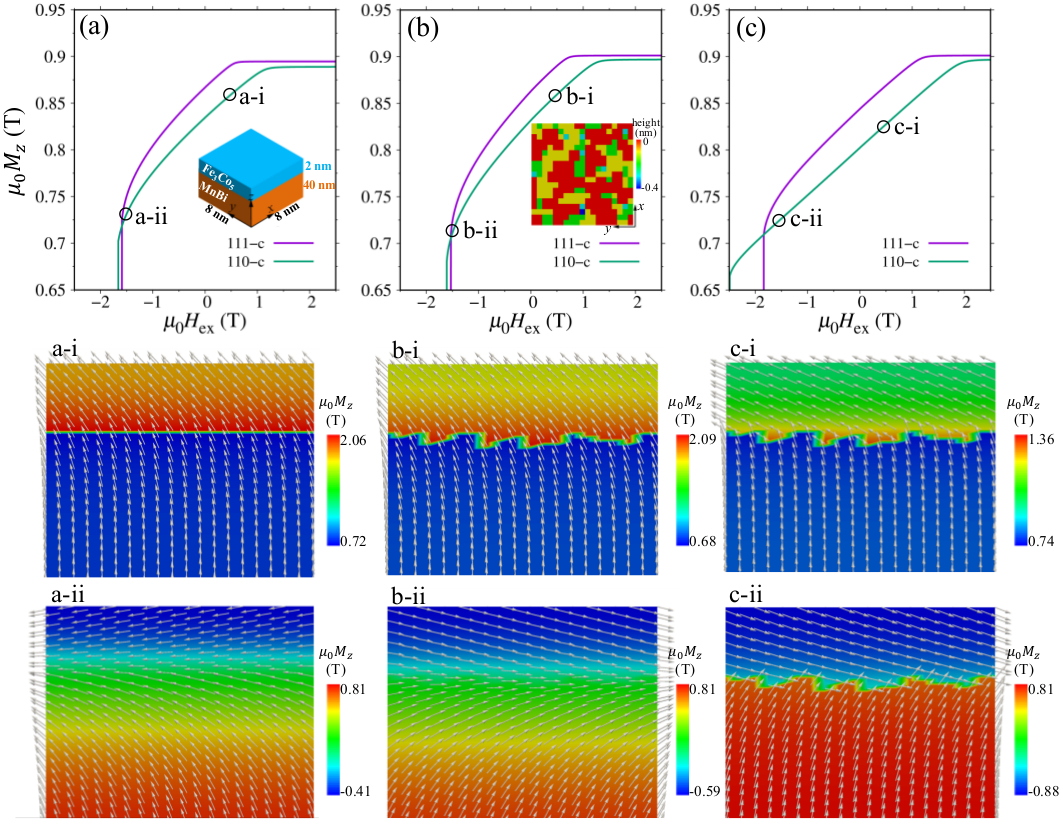}
 	\caption{Micromagnetic simulation results of a MnBi/Fe$_3$Co$_5$ model system with crystalline FeCo(110) and (111) interfaces. Magnetic reversal curves:(a) No interface roughness with the $A^\text{int}$ value listed in Table \ref{t1}; (b) Interface roughness with the same $A^\text{int}$ as in (a); (c) Interface roughness with $A^\text{int}$ reduced to 10$\%$ of that in (a). The external magnetic field $\mu_0H_\text{ex}$ is applied along $z$ direction. Inset of (a): Model geoemetry with in-plane periodic boundary condition. Inset of (b): Interfacial roughness of MnBi with a maximum dent height of 0.4 nm. a-i and a-ii, b-i and b-ii, and c-i and c-ii present the magnetic configurations ($yz$ surface at $x=0$) corresponding to the marked circles of reversal curves in (a), (b), and (c), respectively which belongs to the most favorable case, {\it i.e.} crystalline Fe$_3$Co$_5$(110).}
 	\label{figYi}
 \end{figure*}
Considering the values of lattice misfit and interface formation energy in Tab. \ref{t1}, it is postulated that MnBi(001)/crystalline FeCo(111) interface is slightly less probable to form. 
Moreover,  crystalline Fe$_3$Co$_5$(110) phase has higher values of $J^\text{int}$ and $A^\text{int}$ compared to other configurations which favors a more coherent interfacial exchange coupling. 
However, the coexistence of disordered phases with lower values of $J^\text{int}$ and $A^\text{int}$ has deteriorated the magnetic exchange coupling at the interface in our experimental measurements (see Fig. \ref{MB-FC-M-H}). 

In order to examine the influence of interfacial properties on the exchange coupling behavior, micromagnetic simulations are performed. Since the microstructure of the experimental sample is very complicated and cannot be fully implemented into any simulations, here we  concentrate on a rather simplified model based on single crystalline structures for evaluating the exchange behavior. 

The employed model with an in-plane size 8\,nm$\times$8\,nm, 40\,nm thick MnBi, and 2\,nm thick Fe$_3$Co$_5$ is shown in the inset of Fig.~\ref{figYi}-a. In-plane periodic boundary conditions are applied. Apart from the interface exchange coupling energy the interface roughness as a critical factor, which can influence the interfacial exchange coupling behavior, is evaluated in our micromagnetic simulation analysis (see Fig.~\ref{figYi}).
Here we take the interfaces with disordered Fe$_3$Co$_5$(110) and crystalline Fe$_3$Co$_5$(111) orientations as the model systems for micormagnetic simulations to consider the phases which could be responsible for incoherent exchange coupling observed in magnetic measurements. The following cases are considered:\\
i) Perfect flat interface with the interface exchange stiffness $A^\text{int}_\text{(111)}=4.6$ pJ/m for Fe$_3$Co$_5$(111) orientation and $A^\text{int}_\text{(110)}=1.7$ pJ/m for Fe$_3$Co$_5$(110) orientation, as shown in Fig.~\ref{figYi}-a;\\
ii) Rough interface with a random distribution of dent height (maximum 0.4\,nm, inset of Fig. \ref{figYi}-b) in MnBi and the same values of $A^\text{int}$ in the case i), as shown in Fig.~\ref{figYi}-b;\\
iii) The same rough interface as in the case ii, but with reduced  $A^\text{int}_\text{(111)}=0.46$ pJ/m and $A^\text{int}_\text{(110)}=0.17$ pJ/m, as shown in Fig.~\ref{figYi}-c.\\
The simulated magnetic reversal curves in Fig.~\ref{figYi} do not show the shoulder which was observed in the measured hysteresis loops of the experimental samples. 
As mentioned above, this shoulder is due to the residual in-plane magnetization component of the hard magnetic phase which was not considered in the micromagnetic simulations but rather assuming a full out-of-plane magnetization vector. By increasing the thickness of the soft magnetic layer a reduced rectangularity was observed in the hysteresis loops. Using micromagnetic simulations we examined the magnetic configuration and its evolution around the interface at different external fields, as shown in the second and third rows of Fig.~\ref{figYi}. When the interface is assumed to be perfect and $A^\text{int}_\text{(110)}=1.7$ pJ/m from Tab.~\ref{t1} is used, the magnetization vectors near the interface in FeCo tend to rotate coherently with those in MnBi, as shown in Figs.~\ref{figYi}-a-i and a-ii. This indicates a rather strong interface exchange coupling. When a rough interface was assumed and $A^\text{int}_\text{(110)}$ remained the same, Figs.~\ref{figYi}-b-i and b-ii still suggest strong interface exchange coupling. However, the interface magnetization vectors are much easier to be reversed.
This can be verified by comparing the distribution of the  $z$ component of magnetization ($\mu_0M_z$). For instance, at $\mu_0H_\text{ex}=0.5$ T, the model with rough interface showed a minimum $\mu_0M_z$ ($\mu_0M_z^\text{min}$) of 0.68\,T around the interface (Fig.~\ref{figYi}-b-i), but the model without roughness showed a little higher $\mu_0M_z^\text{min}$ (Fig.~\ref{figYi}-a-i).
The premature reversal in Fig.~\ref{figYi}-b-i and b-ii could be attributed to the local higher demagnetization field induced by the sharp corners or irregularities in the rough interface \cite{oommfyi1,oommfyi2}. Accordingly, the simulated coercivity in Fig.~\ref{figYi}-b was also slightly smaller than that of Fig.~\ref{figYi}-a.
When the interface roughness was assumed to reduce $A^\text{int}_\text{(110)}$ to 0.17\,pJ/m, the magnetic reversal curve were a simple straight line, as shown in Fig.~\ref{figYi}-c.
From the magnetic configurations in Fig.~\ref{figYi}-c-i and c-ii, it can also be found that the magnetization vectors around the interface cross each other and the magnetization in FeCo almost rotates freely, indicating a very poor interface exchange coupling. From Fig.~\ref{figYi} we realize that the interface exchange coupling strength evaluated from DFT calculations of smooth interfaces provides useful insight into the atomic or compositional design of the MnBi/FeCo system. The micromagnetic modeling reveals in addition that the interface roughness and irregular occurrence of defects are also important parameters since it can induce locally premature reversal and, as a consequence, deteriorates the interface exchange coupling.

\begin{figure}[!th]
	\centering
	\includegraphics[width=0.41\textwidth]{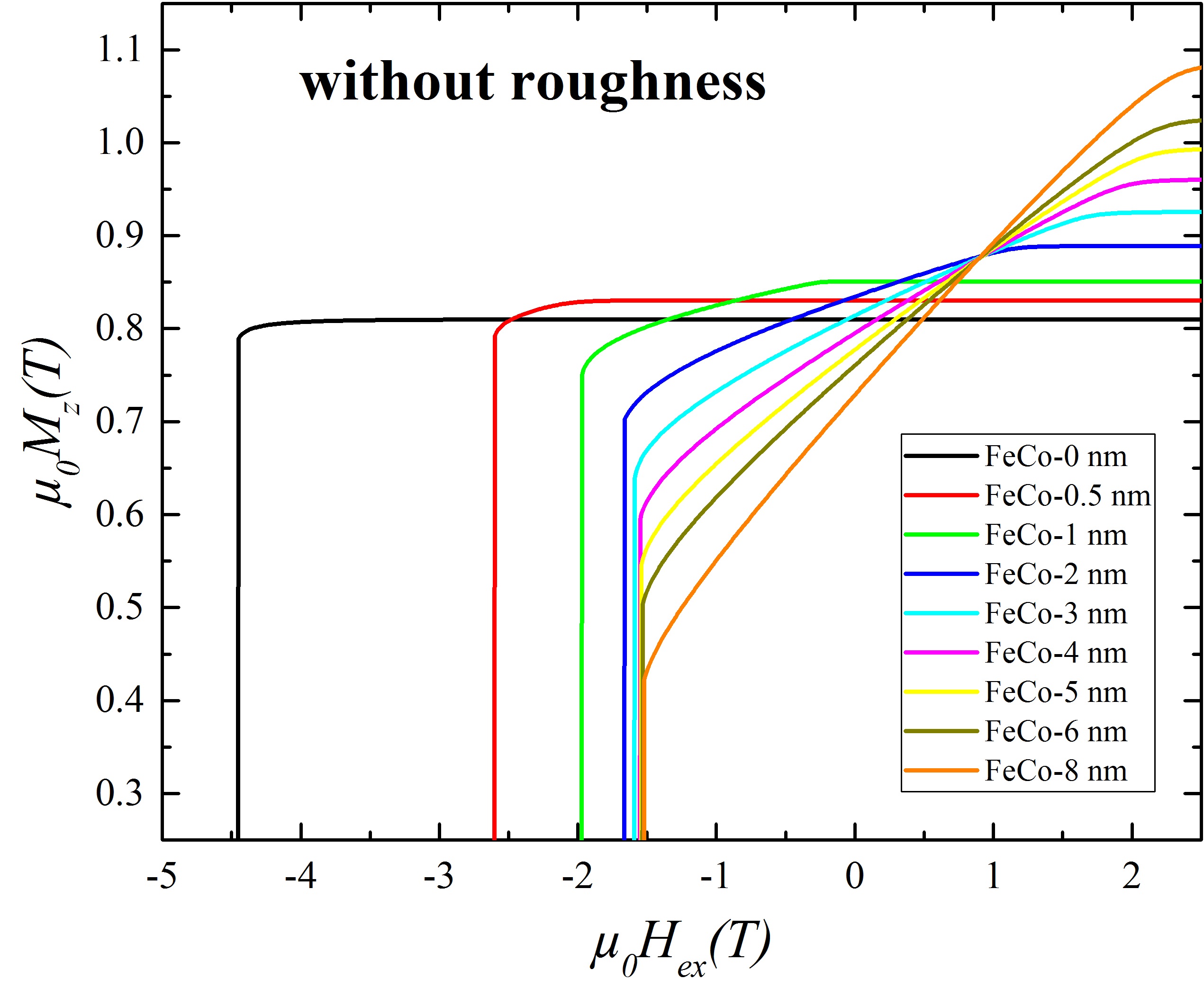}
	\includegraphics[width=0.41\textwidth]{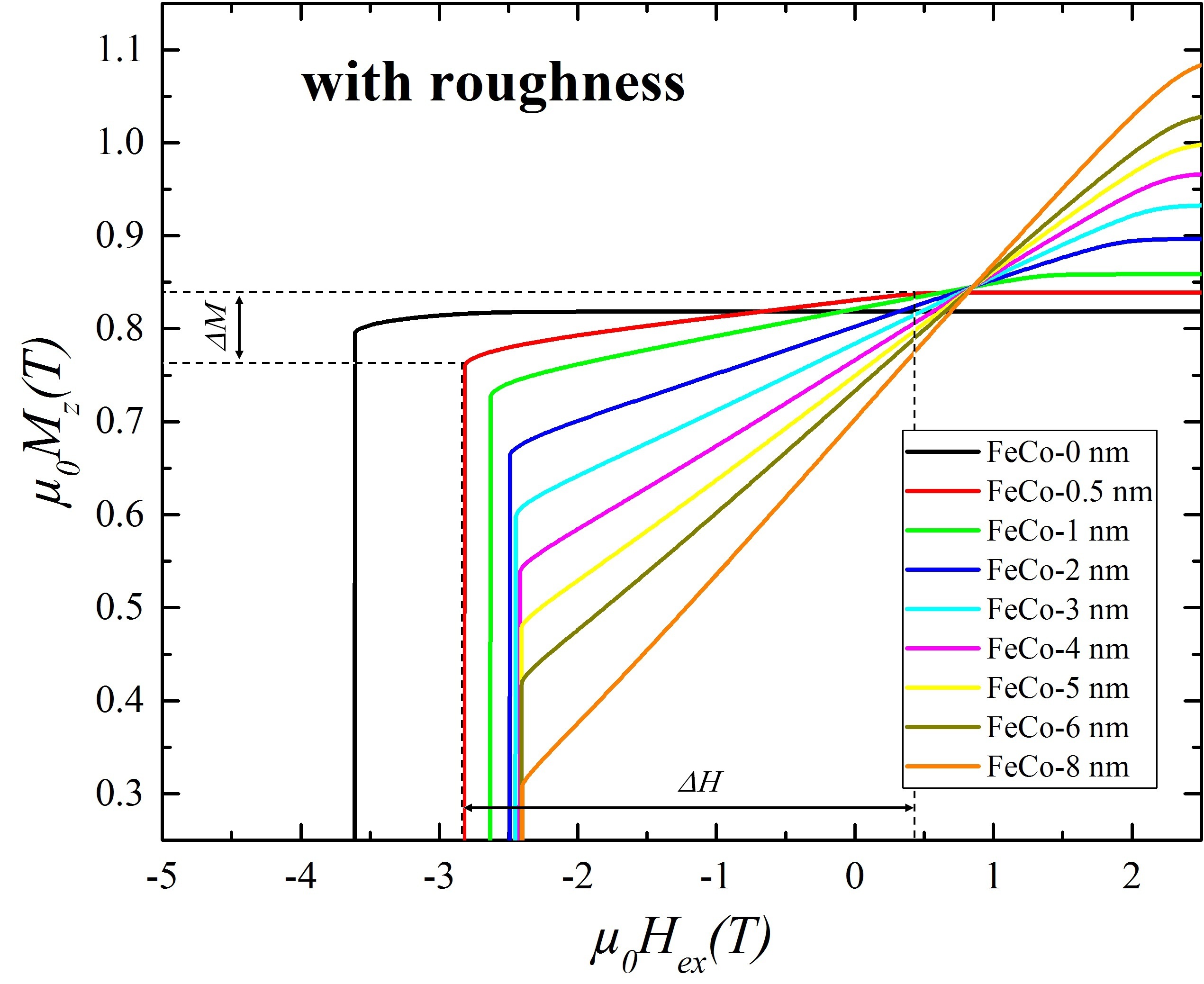}
	\vspace*{0.2cm}
	\includegraphics[width=0.41\textwidth]{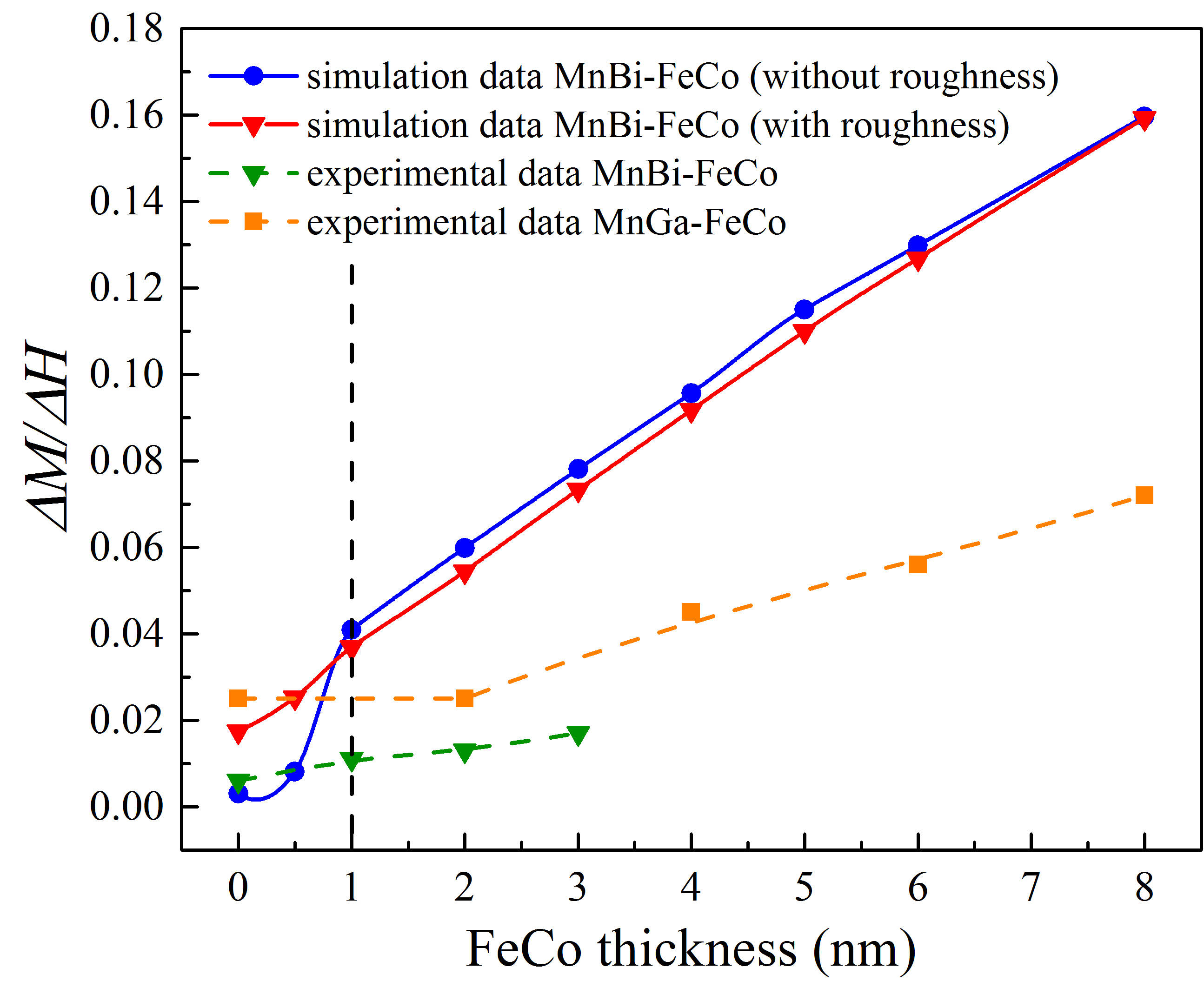}
	\caption{Hysteresis plots obtained from micromagnetic simulations for MnBi(001)/FeCo(110) double layers (a) without and (b) with interface roughness. (c) Variation of the magnetization with respect to the applied field around zero field for the theoretical and experimental hysteresis plots as a function of FeCo thickness. For the case of interfaces without roughness two regions are evident in which at 1\,nm FeCo thickness incoherent coupling between the hard and soft magnetic layers appears. The rough interfaces, in both theoretical and experimental results, behave incoherently from the beginning. For comparison, the experimental data for the epitaxial case of MnGa(001)/FeCo(001) bilayer are also presented.
	}
	\label{origin}
\end{figure}

Fig.~\ref{origin} summarizes the results of thickness analysis based on experimental measurements and theoretical modeling of MnBi/FeCo interface. In Fig.~\ref{origin}-a and -b, 
hysteresis  plots are shown for the two interfaces, namely without and with interface roughness corresponding to the information provided in Fig.~\ref{figYi} for the case of MnBi(001)-disordered Fe$_3$Co$_5$(110).
For each case in Fig.~\ref{origin} the hysteresis is plotted for different thicknesses of the FeCo layer. In the case of a 0.5\,nm thick FeCo layer, it is evident that for the interface without roughness the hysteresis loop is more rectangular (the hysteresis slope at zero-field crossing is close to zero) showing a more coherent coupling compared to the same thickness of FeCo with a rough interface. In order to quantitatively distinguish the changes in exchange coupling considering the effect of interface roughness and increasing the soft layer thickness, the first derivative of the corresponding hysteresis loops has been calculated. The slope around zero-field crossing showing the variation
of magnetization with respect to the applied field ($\frac{\Delta M}{\Delta H}$) for each hysteresis plots as a function of different FeCo thicknesses are shown in Fig.~\ref{origin}-c. 
It should be noted that for the cases of experimental data, the shoulder observed at zero field has been excluded from the derivative plots in order to keep the consistency of the graphs in comparison to the simulation data.
From Fig.~\ref{origin}-c, it can be seen that for structures without interface roughness (blue circle points) two regions are observable. The first region is present up to 1\,nm of FeCo thickness and the second one starts above 1\,nm. It can be seen that for the sample  with less than 1\,nm FeCo thickness without roughness, the hard and soft layers are coherently exchanged coupled since the first derivatives are close to zero. However, considering a rough interface (red triangles with continuous line), exchange coupling is incoherent regardless of the soft layer thickness as the slope is continuously increasing.
Using the same method, the first derivatives of our experimental hysteresis loops corresponding to Fig.~\ref{MB-FC-M-H}-b are plotted as a function of soft layer thicknesses in Fig.~\ref{origin}-c (green triangle with dashed line). In addition, the first derivatives of our experimental magnetization data for epitaxial MnGa(001)/FeCo(001) bilayers have also been included in Fig.~\ref{origin}-c (orange square with dashed line). A complete theoretical and experimental study 
on the MnGa/FeCo exchange spring system is currently under preparation. As can be seen from the plots in Fig.~\ref{origin}-c, our theoretical and experimental findings for the case of MnBi/FeCo bilayer are in agreement and show that the effect of interface roughness on the incoherency of exchange coupling is significant. In addition, it can be concluded that the effect of the lattice misfit between the hard and soft layers is decisive since even in the case of the interfaces without roughness (blue solid line) using a single crystalline model, the coherent coupling is only observed below 1 nm of FeCo thickness. 
These are important findings which provide a better understanding of exchange coupling and go beyond existing knowledge on exchange spring systems.

Comparing the trend of derivative plots for MnBi/FeCo and MnGa/FeCo bilayer systems, it can be seen that as MnGa/FeCo bilayers show much decreased interface roughness due to epitaxial growth, their magnetic data is used here to confirm the micromagnetic modeling approach. The shoulder observed in the slope at zero crossing indicates the transition from coherent to incoherent exchange coupling. While in case of MnBi/FeCo system due to the thin film growth properties only incoherent bilayers were obtained, nevertheless the corresponding graph shows an increasing slope as a function of soft layer thickness which is in agreement with the modeling. In contrast, in the case of MnGa/FeCo epitaxial bilayer a coherent exchange coupling can be obtained up to 2\,nm. Our study shows that not only interface roughness is limiting the interfacial exchange coupling but also
a reduced lattice misfit at interface will greatly improve the coupling behavior.

As a result, finding a suitable single crystal substrate with a small lattice misfit to enable growth of an epitaxial MnBi layer could be one way to improve the exchange coupling behavior in this system. Not only it will result in a better quality of the exchange interface but the total magnetic properties can also be improved by obtaining a higher degree of crystallinity in both hard and soft magnetic layers. Unfortunately, in case of MnBi it is hard to find such single crystalline substrate which matches the crystalline structure and lattice constants of the LTP-MnBi hexagonal phase which makes it very difficult to study the effect of epitaxial growth of MnBi thin films on the exchange coupling in MnBi/FeCo system. Preliminary results on the MnGa/FeCo system show how the combined experimental and theoretical approach described here is of great importance to improve synthesis and performance of future exchange spring material systems.


\section{Summary and conclusion}

In summary, exchange spring $\text{MnBi/Fe}_{x}\text{Co}_{1-x}$ ($x=0.65$ and $0.35$) bilayers with different soft magnetic layer thicknesses were fabricated by DC magnetron sputtering from alloy targets. The magnetic measurements revealed that a Co-rich FeCo soft magnetic layer results in more coherent exchange properties with an optimum soft layer thickness of $\sim$ 1\,nm leading to $\sim$ 3\% increase of the saturation magnetization, however, a complete single-phase hysteresis cannot be obtained for higher FeCo thickness.
A combined theoretical and experimental approach showed that in the MnBi(001)/FeCo system a partially incoherent interface with crystalline and disordered phases is both expected and observed which considerably limits the exchange coupling effect.

As the most important result, micromagnetic simulations showed that the thickness of the soft magnetic layer and the interface roughness between the hard and soft magnetic layers control the effectiveness of exchange coupling. The incomplete exchange coupling observed in MnBi/FeCo bilayers can be correlated with the high interfacial roughness (reducing the exchange constant). Other controlling structural factors include large lattice misfit and coexistence of crystalline and disordered phases  in soft magnetic layer.
Our study suggests that a strong single phase exchange coupling can be extended to higher FeCo thicknesses only through epitaxial growth of both hard and soft magnetic layers with atomically smooth interfaces. Preliminary experimental results show that the MnGa/FeCo system could be a more suited exchange coupling material combination with a critical soft layer thickness of about 2\,nm.


\section{Acknowledgement}

The authors thank the LOEWE project RESPONSE funded by the Ministry of Higher Education, Research and the Arts (HMWK) of the state of Hessen, Germany. We also acknowledge the computer time given by the high performance computer of Hessen ``Lichtenberg'', and K.~Albe from Technische Universit\"at Darmstadt.


\end{document}